\documentclass[journal]{IEEEtran}
\usepackage[utf8]{inputenc}
\usepackage{amsmath,color,graphicx,amssymb,mathtools}
\usepackage{ifthen}
\usepackage{textcomp}
\usepackage{gensymb}
\usepackage{placeins}
\usepackage{float}
\usepackage[bookmarks=false]{hyperref}
\usepackage{tabularx}
\usepackage{epstopdf,multirow}
\usepackage{xcolor}
\usepackage{mathrsfs}
\newcommand{\me}{\mathrm{e}}
\usepackage{tikz,tikz-3dplot}

\newcommand{\rhovec}{\boldsymbol{\rho}}
\newcommand{\rhodvec}{\dot{\rhovec}}
\newcommand{\rhoddvec}{\ddot{\rhovec}}
\newcommand{\rhodddvec}{\dddot{\rhovec}}

\begin{document}
\bstctlcite{IEEEexample:BSTcontrol}
\pagenumbering{gobble}

\title{Orbit Determination Before Detect: Orbital Parameter Matched Filtering for Uncued Detection}

\author{\IEEEauthorblockN{Brendan~Hennessy\IEEEauthorrefmark{1}\IEEEauthorrefmark{2},
Mark~Rutten\IEEEauthorrefmark{3},
Steven~Tingay\IEEEauthorrefmark{1},
Robert~Young\IEEEauthorrefmark{2}
}

\IEEEauthorblockA{\IEEEauthorrefmark{1}International Centre for Radio Astronomy Research, Curtin University, Bentley, WA 6102, Australia}\\
\IEEEauthorblockA{\IEEEauthorrefmark{2}Defence Science and Technology Group, Edinburgh, SA 5111, Australia}\\
\IEEEauthorblockA{\IEEEauthorrefmark{3}InTrack Solutions, Adelaide, SA 5000, Australia}\\   
}

\maketitle
\begin{abstract}	
This paper presents a novel algorithm to incorporate orbital parameters into radar ambiguity function expressions by extending the standard ambiguity function to match Keplerian two-body orbits. A coherent orbital matched-filter will maximise a radar's sensitivity to objects in orbit, as well as provide rapid initial orbit determination from a single detection. This paper then shows how uncued detection searches can be practically achieved by incorporating radar parameters into the orbital search-space, especially for circular orbits. Simulated results are compared to results obtained from ephemeris data, showing that the orbital path determined by the proposed method, and the associated radar parameters that would be observed, match those derived from the ephemeris data.
\end{abstract}

\begin{IEEEkeywords}
space situational awareness; initial orbit determination; radar signal processing; passive radar
\end{IEEEkeywords}

\section{Introduction}
\IEEEPARstart{M}{odern} radar systems are able to generate optimal filters matched to increasingly complex target motion, resulting in increased sensitivity to targets exhibiting these motion at the cost of significant processing load. This problem is most difficult for sensors targeting objects in low Earth orbit (LEO), especially sensors with a significant field of regard. This is due to the observation time required to detect smaller targets, combined with significant orbital velocities and large search volumes, increasing the parameter space to impractical levels.

Extending radar processing integration times in order to increase detection sensitivity requires mitigation against range migration, Doppler migration, and angular migration. The correction of these migrations is further complicated by the motion of the Earth, and hence the sensor located on the Earth. The direct implementation of a matched filter in this radar search space may lead to the incorporation of many parameters.

The nominal trajectory of orbits is well understood and is generally deterministic. The motion of a two-body Keplerian orbit, an idealised case of an object of insignificant mass orbiting around a much larger central body\footnote{Treated as a single point mass.}, can be expressed entirely by six parameters. Matching the processing to this well-defined orbital motion for the purpose of improved radar detection and space situational awareness is therefore a natural extension.

Whilst the primary aim of this general method is to increase a radar's sensitivity to objects in orbit, detections from a filter matched to a target's orbital trajectory will additionally provide coarse initial orbit determination. Traditionally, performing initial orbit determination requires many radar detections of a pass of an object in space.

After briefly covering prior work (\ref{ssec:prior_work}), Section \ref{sec:problem_formulation} details the problem formulation, specifically in terms of ambiguity function expressions (\ref{ssec:ambiguity_surface_generation}) and Keplerian orbital dynamics (\ref{ssec:orbital_dynamics}). In Section \ref{sec:odbd}, Orbit Determination Before Detect (ODBD) methods are discussed, including matched processing to orbital parameters, constraining the search volume (\ref{ssec:odbd_search_constraints}), and constraining the orbit in radar measurement space (\ref{ssec:odbd_zdc}), particularly for uncued detections. Some specific applications, including single-channel object detection and orbit determination are also discussed (\ref{ssec:odbd_single_channel}). Section \ref{sec:results} presents simulated results, with comparison against ephemerides. Section \ref{sec:conclusion} concludes with a description of future work.
\vspace{-1ex}

\subsection{Prior Work}
\label{ssec:prior_work}
The motivation for this paper is to further develop techniques for the surveillance of space with the Murchison Widefield Array (MWA) using passive radar. The paper is particularly concerned with developing techniques for uncued detection over a wide field of regard. The MWA is a low frequency (70 - 300 MHz), wide field-of-view, radio telescope located in Western Australia \cite{2013PASA...30....7T}. The MWA has demonstrated the incoherent detection of the International Space Station (ISS) \cite{ 2013AJ....146..103T} and other, smaller, objects in orbit \cite{prabu2020development}. However, for coherent processing, methods compensating for all aspects of motion migration are required in order to detect smaller satellites and space debris \cite{8835821}. As passive radar systems have no control over the transmitter used for detection, improving processing gain through extended Coherent Processing Intervals (CPIs) is a method used to achieve the required sensitivity \cite{4653940}. Orbital trajectories are ideal targets for such techniques, as stable and predictable relative motion allows for simpler measurement models. Such techniques have also been used with active radar, for improved sensitivity and processing gain \cite{markkanen2005real} \cite{8812975}.

Consisting of 256 tiles spread across many square kilometres, the MWA's sparse layout\footnote{At FM radio frequencies, even the compact configuration of MWA Phase II is sparse \cite{pase22018article}.} provides high angular resolution. Objects in orbit will therefore transit many beamwidths per second at the point of closest approach.  Because of this, high angular resolution (normally a desirable attribute) can result in significant angular migration. Highly eccentric orbits will transit significantly faster. This is particularly challenging for the uncued detection of small objects, where longer integration times are needed to achieve sufficient sensitivity.

Individual radar detections consisting of a single measurement of range, Doppler, azimuth and elevation, only define a broad region of potential orbital parameters \cite{2007CeMDA..97..289T}. This region may be constrained by incorporating angular rates \cite{2014demarsradar_regions1}, and even further by including radial acceleration and jerk \cite{8812975}.  Usually,  many radar detections are required to perform initial orbit determination. The mapping between radar measurement space and orbital parameters is an ongoing area of research \cite{8448187}.

\section{Problem Formulation}
\label{sec:problem_formulation}
\subsection{Radar Product Formation} 
\label{ssec:ambiguity_surface_generation}
A standard timeseries matched filter is a function to detect reflected copies of a reference signal $d(t)$ in the surveillance signal $s(t)$, specifically copies delayed by $\tau$ and frequency shifted by $f_D$:
\vspace{-1ex}
\begin{IEEEeqnarray}{rCL}
    \label{eq:woodward1}
    \chi(\tau,f_D) = \int_T s(t){d}^*(t-\tau){\me}^{-j2\pi f_Dt}\,dt .
\end{IEEEeqnarray}

This matched filter can be extended to more complicated motions by \textit{dechirping} (or even applying higher order corrections to) the motion-induced frequency shift. For example, instead of matching to the radial velocity with a Doppler shift of $f_D$, higher order motions could be matched with a time varying frequency (that can be represented as a polynomial phase signal) given by $f_D +
f_Ct$, where $f_C$ is proportional to the radial acceleration. This can be extended to an arbitrary number of parameters at the cost of adding extra dimensions to the matched filter outputs. 
To account for any range migration, the delay term $\tau$ will also need to be a function of time to match the radial motion.

For a receiver array consisting of $N$ elements, the surveillance signal $s(t)$ can be formed by classical far-field beamforming in a direction of interest such that:
\begin{IEEEeqnarray}{rCL}
s(t) = \sum_{n=1}^{N}s_{n}(t){\me}^{-j\boldsymbol{k}(\theta,\phi) \cdot \boldsymbol{u}_n } ,
\end{IEEEeqnarray}
where $s_{n}(t)$ is the received signal at the $n^{\text{th}}$ antenna, $\boldsymbol{u}_n$ is the position of the $n^{\text{th}}$ antenna, and $\boldsymbol{k}(\theta,\phi)$ is the signal wavevector for azimuth $\theta$ and elevation $\phi$. 
Time varying adjustments can be made to every measurement parameter to create a filter, $\chi$, matched to the exact motion of an object with range $\rho(t)$ and slant range-rate $\dot{\rho}(t)$, in time-varying directions given by azimuth $\theta(t)$ and elevation $\phi(t)$:

\begin{multline}
  \label{eq:time_varying_delay_doppler}
  \chi\left(\theta(t), \phi(t), \rho(t), \dot{\rho}(t)\right) = \int_T \left[\sum_{n=1}^{N}{\me}^{j\boldsymbol{k}(\theta(t),\phi(t)) \cdot \boldsymbol{u}_n }s_{n}(t)\right] \\
  d^*\left(t-2c^{-1}\rho(t)\right){\me}^{-j\frac{4\pi}{\lambda}\dot{\rho}(t)t}dt ,
\end{multline}
where the delay to the target is now given by the total path distance scaled by $\frac{1}{c}$, and the Doppler shift is given by $\frac{2\dot{\rho}}{\lambda}$.

\subsection{Orbital Dynamics}
\label{ssec:orbital_dynamics}
\vspace{-.47ex}
The most common elements used to parameterise an orbit are the Keplerian, or \textit{classical}, orbital elements. These elements directly describe the size, shape, and orientation of an orbital ellipse (with one focus being at the centre of the central body), and the position of an object on this ellipse at some epoch, in the Earth-Centered Inertial (ECI) coordinate frame \cite{Vallado2001fundamentals}. The ECI coordinate frame has its origin at the centre of the Earth, but it does not rotate with the Earth. It is also worth noting that a Keplerian orbit can, in fact, be any conic section. However, in this paper, it is assumed that orbits describe Earth-captured closed orbits.

\begin{figure}[ht!]
\hspace{2ex}
\tdplotsetmaincoords{70}{110}
\begin{tikzpicture}[tdplot_main_coords,scale=4.3]
  \pgfmathsetmacro{\r}{0.7}
  \pgfmathsetmacro{\O}{45} 
  \pgfmathsetmacro{\i}{35} 
  \pgfmathsetmacro{\f}{32}

  \coordinate (O) at (0,0,0);

  \draw [->] (O) -- (1,0,0) node[anchor=north east] {$\boldsymbol{I}$};
  \draw [->] (O) -- (0,0.9,0) node[anchor=north west] {$\boldsymbol{J}$};
  \draw [->] (O) -- (0,0,0.9) node[anchor=south] {$\boldsymbol{K}$};
  
  \tdplotdrawarc[dashed]{(O)}{\r}{0}{360}{}{}

  \tdplotsetrotatedcoords{\O}{0}{0}

  \draw [tdplot_rotated_coords] (0,0,0) -- (\r,0,0) node [below right] {};
  \tdplotdrawarc[->]{(O)}{.28*\r}{0}{\O}{anchor=north}{$\Omega$}

  \tdplotsetrotatedcoords{-\O}{\i}{0}
  \tdplotdrawarc[tdplot_rotated_coords]{(O)}{\r}{0}{360}{}{}  
  \begin{scope}[tdplot_rotated_coords]
    \draw[->] (O) -- (0,0,0.8) node [above] {$\boldsymbol{h}$};
    \draw (0,0,0) -- (-\r,0,0);
    \tdplotdrawarc[->]{(O)}{.4*\r}{90}{180}{anchor=west}{$\omega$}
    \coordinate (P) at (180+\f:\r);
    \draw (O) -- (P);
    \tdplotdrawarc[->]{(O)}{.7*\r}{180}{180+\f}{anchor=south west}{$\nu$}
  \end{scope}

  \tdplotsetrotatedcoords{-\O+\f}{\i}{0}
  \tdplotsetrotatedcoordsorigin{(P)}
  \begin{scope}[tdplot_rotated_coords,scale=.2,thick]
    \fill (P) circle (.6ex) node [above] {Celestial Body};
  \end{scope}

  \tdplotsetthetaplanecoords{-\f}
  \tdplotdrawarc[tdplot_rotated_coords,->]{(O)}{0.9*\r}{0}{\i}{anchor=south}{$i$}
\end{tikzpicture}
\vspace{-2ex}
\caption{The orbital plane determined by orientation parameters $\Omega$, $\omega$, and $i$ relative to the plane of reference in the ECI coordinate frame. These parameters define the direction of the angular momentum vector $\boldsymbol{h}$. The axes $\boldsymbol{I}$, $\boldsymbol{J}$ and $\boldsymbol{K}$ define the ECI coordinate frame.}
\vspace{-1ex}
\label{fig:orbitplane1}
\end{figure}
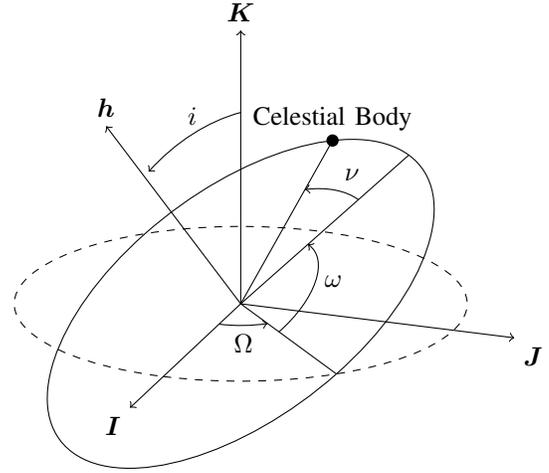

The Keplerian orbital parameters are: the semi-major axis, $a$, and eccentricity, $e$, defining the size and shape of the ellipse; the right-ascension of the ascending node, $\Omega$, and inclination, $i$, which define the orientation of the elliptical plane to the Earth's equatorial plane; the argument of periapsis, $w$, defining the orientation/rotation of the ellipse in the orbital plane; and finally, the true anomaly, $\nu$, defining the position of the object on the ellipse (refer to Figure \ref{fig:orbitplane1}).

\vspace{-3ex}

It is also assumed that the only force acting on the object in orbit is due to the gravity of the dominant mass\footnote{Uniform acceleration does not take into account the ellipsoidal/oblate nature of the Earth or other forces, such as micro-atmospheric drag, solar weather, and gravity due to other celestial bodies. For the short duration of a single CPI, these factors are generally negligible.}, with the acceleration due to the Earth's gravity $\boldsymbol{\ddot{r}}$, given by:
\vspace{-0.2ex}
\begin{equation}
\boldsymbol{\ddot{r}}= -\frac{\mu}{\lvert\boldsymbol{r}\rvert^3}\boldsymbol{r}\label{eq:eci_acceleration},\vspace{-1ex}
\end{equation}
where $\mu$ is the standard gravitational parameter for the Earth.

Given the orbital parameters, and the acceleration due to the Earth's gravity, the Cartesian position $\boldsymbol{r}$, and velocity $\boldsymbol{\dot{r}}$, for an object in Earth orbit is completely deterministic and is given by:
\begin{IEEEeqnarray}{rCL}
    \boldsymbol{r} &=& \frac{a(1-e^2)}{1 + e\cos{\nu}}(\cos{\nu}\boldsymbol{P} + \sin{\nu}\boldsymbol{Q})\label{eq:eci_position}~;\\
    \boldsymbol{\dot{r}} &=& \sqrt{\frac{\mu}{a(1-e^2)}}(-\sin{\nu}\boldsymbol{P} + (e + \cos{\nu})\boldsymbol{Q})\label{eq:eci_velocity}~,
\end{IEEEeqnarray}
where $\boldsymbol{P}$ and $\boldsymbol{Q}$ represent axes of a coordinate system co-planar with the orbital plane in the Cartesian ECI coordinate frame (given by axes $\boldsymbol{I}$, $\boldsymbol{J}$, and  $\boldsymbol{K}$). The third axis, $\boldsymbol{W}$, is perpendicular to the orbital plane \cite{Vallado2001fundamentals}. These vectors are described by:

\begin{IEEEeqnarray}{rCL}
\boldsymbol{P} = & &
    \begin{bmatrix} 
    \cos{\Omega}\cos{\omega} - \sin{\Omega}\cos{i}\sin{\omega} \\
    \sin{\Omega}\cos{\omega} + \cos{\Omega}\cos{i}\sin{\omega} \\
    \sin{i}\sin{\omega}
    \end{bmatrix}~; \\
    \boldsymbol{Q} = & & \begin{bmatrix} 
    -\cos{\Omega}\sin{\omega} - \sin{\Omega}\cos{i}\cos{\omega} \\
    -\sin{\Omega}\sin{\omega} + \cos{\Omega}\cos{i}\cos{\omega} \\
    \sin{i}\cos{\omega}
    \end{bmatrix}~;\\
    \boldsymbol{W} = & & \begin{bmatrix} 
    \sin{i}\sin{\Omega} \\
    -\sin{i}\cos{\Omega} \\
    \cos{i}
    \end{bmatrix}~.
\end{IEEEeqnarray}
Note that a complicating factor with the ECI reference frame is that a nominally stationary position on the surface of the Earth, such as a fixed radar sensor, will have significant motion.

\section{{Orbit Determination Before Detect}} \label{sec:odbd}

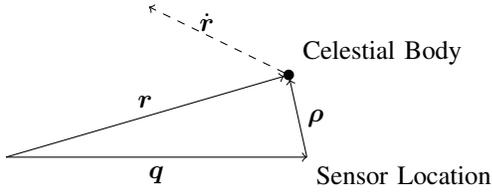
\begin{figure}[ht!]
\hspace{7ex}
\begin{tikzpicture}[dot/.style={draw,fill,circle,inner sep=1pt}]
  \def\a{5}
  \def\b{\a}
  \def\angle{10} 
  \coordinate[] (O) at (0,0) {};
  \node[thick, dot,label={\angle:Celestial Body}] (X) at (20:{4} and {3.2}) {};
  \coordinate[label={-30:Sensor Location}] (Q) at (0:{4} and {3}) {};
  \draw [->] (O) -- (X) node [midway, above] {$\boldsymbol{r}$};
  \draw [->] (O) -- (Q) node [midway, below] {$\boldsymbol{q}$};
  \draw [dashed,  ->] (X) -- (1.9,2) node [near end, right] [shift=(10:1mm)] {$\dot{\boldsymbol{r}}$};
  \draw [ ->] (Q) -- (X) node [midway, right] {$\boldsymbol{\rho}$};
  \fill (X) circle [radius=2pt];

\end{tikzpicture}
\vspace{-2ex}
\caption{In the ECI coordinate frame the sensor is at position $\boldsymbol{q}$, the celestial body at position $\boldsymbol{r}$ with velocity $\dot{\boldsymbol{r}}$ (given by \eqref{eq:eci_position} and \eqref{eq:eci_velocity}) and the slant range vector from the sensor to the object given by $\boldsymbol{\rho}$.}
\label{fig:radar_vectors}
\end{figure}

For a two-body Keplerian orbit, the time-varying terms $\rho(t)$, $\dot{\rho}(t)$, $\phi(t)$, and $\theta(t)$ 
\eqref{eq:time_varying_delay_doppler} can be completely described by an orbit's six independent parameters. Although the position of an object in orbit is given by \eqref{eq:eci_position}, there are no closed form solutions for the time varying position $\boldsymbol{r}(t)$. Instead, a Taylor series approximation can be used to calculate an expression for the object's position throughout a CPI such that $\boldsymbol{r}(t) = \sum_{n=0}^{\infty}\frac{\boldsymbol{r}^{(n)}(0)t^n}{n!}$ (where $\boldsymbol{r}^{(n)}(x)$ denotes the $n^{th}$ derivative of $\boldsymbol{r}$ evaluated at the point $x$), with $t$ being the time through the CPI of length $T$, $t \in [\frac{-T}{2}, \frac{T}{2}]$. With knowledge of the sensor's location, $\boldsymbol{q}(t)$ (as in Figure \ref{fig:radar_vectors}), and $\boldsymbol{\dot{q}}(t)$ giving the slant range vector from the sensor to the object, as well as the slant-range rate, as $\boldsymbol{\rho}(t) = \boldsymbol{r}(t) - \boldsymbol{q}(t)$ and $\boldsymbol{\dot{\rho}}(t) = \boldsymbol{\dot{r}}(t) - \boldsymbol{\dot{q}}(t)$, a polynomial expression for the slant-range and slant-range rate equations of motion over the CPI is possible:
\vspace{-1ex}
\begin{IEEEeqnarray}{rCL}
    \rho(t) &=& \lvert\boldsymbol{\rho}(t)\rvert = \lvert \sum_{n=0}^{\infty}\frac{\boldsymbol{r}^{(n)}(0)t^n}{n!} - \boldsymbol{q}(t)\rvert\label{eq:taylor_position}~;\\
     \dot{\rho}(t) &=& \lvert\boldsymbol{\dot{\rho}}(t)\rvert = \lvert \sum_{n=1}^{\infty}\frac{\boldsymbol{r}^{(n)}(0)t^n}{n!} - \boldsymbol{\dot{q}}(t)\rvert\label{eq:taylor_rangerate}~.
\end{IEEEeqnarray}
These expressions can be extended (or truncated) to arbitrary accuracy.

The directional angles are now calculated as topocentric right ascension and declination, that is right ascension and declination relative to the sensor location, given by $\alpha$ and $\delta$, respectively: 
\vspace{-1.5ex}
\begin{IEEEeqnarray}{rCL}
    \alpha(t) &=& {\tan}^{-1}\left(\frac{\rho_{\boldsymbol{J}}(t)}{\rho_{\boldsymbol{I}}(t)}\right)\label{eq:alpha_t}~;\\
    \delta(t) &=& {\tan}^{-1}\left(\frac{\rho_{\boldsymbol{K}}(t)}{\sqrt{{\rho_{\boldsymbol{I}}(t)}^2 + {\rho_{\boldsymbol{J}}(t)}^2}}\right)~, \label{eq:delta_t}
\end{IEEEeqnarray}
noting that these expressions depend on the individual elements of $\boldsymbol{\rho}$ such that $\boldsymbol{\rho}(t) = [\rho_{\boldsymbol{I}}(t), \rho_{\boldsymbol{J}}(t), \rho_{\boldsymbol{K}}(t)]^T$.

Using the expressions in this section, it is possible to form a matched filter to the orbital elements themselves, essentially creating $\chi(e,a,i,\Omega,\omega,\nu)$ at a given epoch \eqref{eq:time_varying_delay_doppler}. This enables arbitrarily long CPIs by tracking an orbit throughout the CPI. Additionally, instead of calculating a Taylor Series expression for the orbital position $\boldsymbol{r}(t)$, and deriving the parameters of interest, it is far more efficient to directly calculate a Taylor Series expression for the parameters of interest. 
For a sensor at known Cartesian position $\boldsymbol{q}$, with known instantaneous velocity, acceleration and jerk, given by $\boldsymbol{\dot{q}}$, $\boldsymbol{\ddot{q}}$, and $\boldsymbol{\dddot{q}}$, respectively, and given the slant range vector $\boldsymbol{\rho} = \boldsymbol{r} - \boldsymbol{q}$, the slant range and its instantaneous derivatives are given by:

\begin{align}
  \rho &= \lvert\rhovec\rvert\label{eq:straight_up_slant_range}~;\\
  \label{eq:orbit_doppler}
  \dot{\rho} &= \frac{\rhovec\cdot\rhodvec}{\rho}~; \\
  \label{eq:orbit_chirp}
  \ddot{\rho} &= -\frac{(\rhovec\cdot\rhodvec)^2}{\rho^3} +  \frac{|\rhodvec|^2 + \rhovec\cdot\rhoddvec}{\rho}~; \\
  \label{eq:orbit_jerk}
  \dddot{\rho} &= \begin{multlined}[t]
     3\frac{(\rhovec\cdot\rhodvec)^3}{\rho^5} \\
     - 3\frac{(\rhovec\cdot\rhodvec)(|\rhodvec|^2 + \rhovec\cdot\rhoddvec)}{\rho^3}\\
     + \frac{3\rhodvec\cdot\rhoddvec + \rhovec\cdot\rhodddvec}{\rho}~,
  \end{multlined}
\end{align}

where $\dddot{\boldsymbol{r}}$ is from the derivative of \eqref{eq:eci_acceleration} and is given by:
\begin{IEEEeqnarray}{rCL}
\dddot{\boldsymbol{r}} =  \frac{3\mu\boldsymbol{r}\cdot\boldsymbol{\dot{r}}}{\lvert\boldsymbol{r}\rvert^5}\boldsymbol{{r}} -\frac{\mu}{\lvert\boldsymbol{r}\rvert^3}\boldsymbol{\dot{r}}~.
\end{IEEEeqnarray}

Now, \eqref{eq:orbit_doppler}, \eqref{eq:orbit_chirp}, and \eqref{eq:orbit_jerk} can be used to directly specify the target's Doppler, chirp rate, and radial jerk. This leads to more efficient expressions (when compared to 
\eqref{eq:taylor_position} and \eqref{eq:taylor_rangerate}) for the slant-range, and also slant-range rate, throughout the CPI of length $T$ such that $t \in [\frac{-T}{2}, \frac{T}{2}]$: 
\begin{IEEEeqnarray}{rCL}
    \rho(t) & = & \rho + \dot{\rho}t + \frac{1}{2}\ddot{\rho}t^2 + \frac{1}{6}\dddot{\rho}t^3~; \\
    \dot{\rho}(t) & = & \dot{\rho} + \ddot{\rho}t + \frac{1}{2}\dddot{\rho}t^2~.
\end{IEEEeqnarray}

A fourth-order Taylor Series approximation to the slant-range, $\rho(t)$, was chosen due to previous work, which demonstrated that a third order polynomial phase signal may be required in order to coherently match orbits for CPIs of duration up to 10 seconds \cite{8835821}. 

Similarly, equivalent approximations can be formed for the angular measurement parameters $\alpha(t)$ 
\eqref{eq:alpha_t} and $\delta(t)$ \eqref{eq:delta_t}.
\subsection{Search-Volume Constraints}
\label{ssec:odbd_search_constraints}
The methods described above enable coherent processing that matches orbital parameters; however, they are not suitable for searching to perform uncued detections. The parameter space is far too large to be practically searched, and the vast majority of orbits will not correspond to passes within a region of interest above the sensor. Although, as stated earlier in Section \ref{ssec:orbital_dynamics}, alternatives to the Keplerian parameter set are available. In fact, it is possible to parameterise a Keplerian orbit with the Cartesian position and velocity to constitute the six elements \cite{Vallado2001fundamentals}. It is also possible to utilise combinations of both sets of elements in other formulations.

Instead of searching through classical orbital parameters, three parameters can be expressed as a hypothesised ECI position within a search volume of interest. This ensures any hypothesised orbit, determined from these initial parameters, will be within the search volume. Given this potential orbital position, $\boldsymbol{r}$, only three more additional parameters are needed to fully define an elliptical orbit. Although the three elements forming the orbital velocity could be treated as free variables, the majority of possible velocities would not correspond to valid Earth-captured orbits. Instead, given position $\boldsymbol{r}$ and semi-major axis $a$, the magnitude of the velocity of the corresponding orbit is given by the Vis-Viva equation \cite{Vallado2001fundamentals}:
\begin{IEEEeqnarray}{rCL}
{\lvert\boldsymbol{\dot{r}}\rvert}^2
= \mu(\frac{2}{\lvert\boldsymbol{r}\rvert} - \frac{1}{a})~. \label{eq:vis_viva}
\end{IEEEeqnarray}

Furthermore, given position $\boldsymbol{r}$ and eccentricity $e$, the semi major axis length will itself be constrained between the potential limits of the orbit's apogee and perigee ranges:
\begin{IEEEeqnarray}{rCL}
\frac{\lvert\boldsymbol{r}\rvert}{1+e} \leq a \leq \frac{\lvert\boldsymbol{r}\rvert}{1-e}~. \label{eq:apogiee_and_perigee_ranges}
\end{IEEEeqnarray}

The semi-major axis is also constrained by realistic limits on an orbit's range, as well as a sensor's maximum detection range, represented by minimum and maximum allowable periapsides, ${rp}_{min}$ and ${rp}_{ max}$:
\begin{IEEEeqnarray}{rCL}
\frac{{rp}_{min}}{1-e} \leq a \leq \frac{{rp}_{max}}{1-e}~. \label{eq:perigee_ranges_limits}
\end{IEEEeqnarray}

Another constraint is the constant angular momentum of the orbit, $\boldsymbol{h}$. This vector is perpendicular to the orbital plane, parallel to $\boldsymbol{W}$\hspace{-0.5ex}, with a magnitude depending on the size and shape of the ellipse:
\begin{IEEEeqnarray}{rCL}
    \boldsymbol{h} = \sqrt{\mu a(1-e^2)}\boldsymbol{W} = \boldsymbol{r}\times\boldsymbol{\dot{r}}~. \label{eq:angular_momentum}
\end{IEEEeqnarray}

This cross-product may be rewritten to form an expression for the inner product between the position and velocity:
\begin{IEEEeqnarray}{rCL}
    \boldsymbol{r}\cdot\boldsymbol{\dot{r}} & = & \pm\sqrt{\lvert\boldsymbol{r}\rvert^2\lvert\boldsymbol{\dot{r}}\rvert^2 - \lvert\boldsymbol{h}\rvert^2}~.
\end{IEEEeqnarray}

Combined with the magnitude of the velocity, from the Vis-Viva equation \eqref{eq:vis_viva}, as well as the magnitude of the constant angular momentum \eqref{eq:angular_momentum}, an expression for this inner product can be formed which depends solely on the position $\boldsymbol{r}$ and the size and shape of the orbital ellipse:
\begin{IEEEeqnarray}{rCL}
    \boldsymbol{r}\cdot\boldsymbol{\dot{r}} & = & \pm\sqrt{\lvert\boldsymbol{r}\rvert^2\mu(\frac{2}{\lvert\boldsymbol{r}\rvert} - \frac{1}{a}) - \mu a (1-e^2)}~. \label{eq:parallel_planes}
\end{IEEEeqnarray}

Additionally, the specific relative angular momentum vector, $\boldsymbol{h}$, is perpendicular to both the orbital position $\boldsymbol{r}$ and orbital velocity $\boldsymbol{\dot{r}}$. This leads to the expressions $\boldsymbol{r}\cdot\boldsymbol{h}=0$ and $\boldsymbol{\dot{r}}\cdot\boldsymbol{h}=0$, which result in another constraint on the velocity, dependant on the right ascension of the ascending node, $\Omega$:
\begin{IEEEeqnarray}{rCL}
    \begin{bmatrix}
         r_{\boldsymbol{K}}\sin{\Omega} \\ -r_{\boldsymbol{K}}\cos{\Omega} \\ r_{\boldsymbol{J}}\cos{\Omega} - r_{\boldsymbol{I}}\sin{\Omega}
        \end{bmatrix}\cdot\boldsymbol{\dot{r}} = 0~. \label{eq:raan_plane}
\end{IEEEeqnarray}

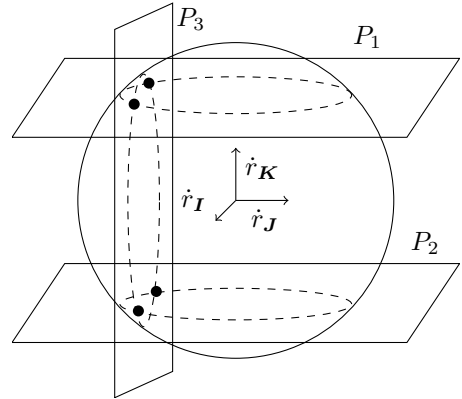
\begin{figure}[ht!]
\hspace{8ex}\begin{tikzpicture}[
  point/.style = {draw, circle, fill=black, inner sep=0.7pt},
  scale=0.7
]\clip (-4.25,-3.75) rectangle + (8.5,8);
\def\rad{3cm}
\coordinate (O) at (0,0); 

  \draw[->] (0,0,0) -- (1,0,0) node[anchor=north east]{$\dot{{r}}_{\boldsymbol{J}}$};
  \draw[->] (0,0,0) -- (0,1,0) node[anchor=north west]{$\dot{{r}}_{\boldsymbol{K}}$};
  \draw[->] (0,0,0) -- (0,0,1) node[anchor=south east]{$\dot{{r}}_{\boldsymbol{I}}$};

\draw[-] (0,0) circle [radius=\rad];

\begin{scope}[]
\draw[-]
  (-3.25,2.7) -- (4.25,2.7) -- (3.25,1.2) -- (-4.25,1.2) -- cycle;
  
  \draw[]
  (-3.25,-1.2) -- (4.25,-1.2) -- (3.25,-2.7) -- (-4.25,-2.7) -- cycle;
  
  \draw[]
    (-2.3,3.25) -- (-2.3,-3.75) -- (-1.2,-3.25) -- (-1.2,3.75) -- cycle;
  
  \draw[dashed] (0,2) ellipse (2.2 and 0.35);
  \draw[dashed] (0,-1.97) ellipse (2.2 and 0.3);
  \draw[dashed] (-1.75,0) ellipse (0.3 and 2.4 );
  
\node at (2.5,3.1) {$P_1$};

\node at (3.6,-0.8) {$P_2$};

\node at (-0.85,3.45) {$P_3$};

\fill (-1.65,2.23) circle [radius=0.105];

\fill (-1.93,1.83) circle [radius=0.105];

\fill (-1.51,-1.73) circle [radius=0.105];

\fill (-1.85,-2.1) circle [radius=0.105];

\end{scope}
\end{tikzpicture}
\vspace{-1.5ex}
\caption{Four valid orbital velocities given by the intersection of the sphere (given by \eqref{eq:vis_viva}), parallel planes P1 and P2 (given by \eqref{eq:parallel_planes}), and plane P3 (given by \eqref{eq:raan_plane} or \eqref{eq:doppler_plane}).}
\label{fig:makingshapes}
\end{figure}

These expressions lead to a simple geometric solution for determining orbits when $\boldsymbol{r}$ (and other parameters) are known, and $\boldsymbol{\dot{r}}$ is unknown. For determining $\boldsymbol{\dot{r}}$, 
(\ref{eq:vis_viva}) defines a  sphere of radius $\sqrt{\mu(\frac{2}{\lvert\boldsymbol{r}\rvert} - \frac{1}{a})}$, representing valid orbits in the velocity vector's element space. Additionally, 
(\ref{eq:parallel_planes}) defines two parallel planes of valid orbits, which intersect with (\ref{eq:vis_viva}) to define two circles. Finally, intersecting these two circles with the plane defined by the position and the right ascension of the ascending node, $\Omega$, 
\eqref{eq:raan_plane} will result in a maximum of four intersection points, that is, four velocities, each corresponding to a valid orbit. An example diagram is shown in Figure \ref{fig:makingshapes}. Although this means that a choice of six orbital parameters will result in up to four potential orbital matched filters, this approach will be far more efficient than methods outlined earlier in this section, as the orbit will be within the search volume, and each parameter choice restricts the range of subsequent parameters. 

\vspace{-3ex}

Therefore, given an orbital position, $\boldsymbol{r}$, a choice of eccentricity, $e$, semi-major axis, $a$, and right ascension of the ascending node, $\Omega$, four potential orbital velocities, $\boldsymbol{\dot{r}}$, are calculated, which leads to an expression for the complete matched filter:
\vspace{-0.5ex}
\begin{IEEEeqnarray}{rCL}
    \chi(\boldsymbol{r}, \boldsymbol{\dot{r}}) 
        & = \hspace{-1ex} \int\limits_{-\frac{T}{2}}^{\frac{T}{2}} \hspace{-1ex} [&\sum_{n=1}^{N}{\me}^{j\boldsymbol{k}(\delta(\boldsymbol{r}, \boldsymbol{\dot{r}},t),\alpha(\boldsymbol{r}, \boldsymbol{\dot{r}},t)) \cdot \boldsymbol{u}_n }s_{n}(t)]\nonumber\\
     & & ~d^*(t-2c^{-1}\rho(\boldsymbol{r}, \boldsymbol{\dot{r}},t)){\me}^{-j\frac{2\pi}{\lambda}\dot{\rho}(\boldsymbol{r}, \boldsymbol{\dot{r}},t)t}\,dt~.~~~\label{eq:full_OD_right_here}
\end{IEEEeqnarray}

The proposed method tests for only realistic orbits in a given search region. Also, given a set of orbit parameters, this matched filter should maximise a radar's sensitivity to that orbit. Additionally, a detection in this matched filter corresponds to a detection in the orbital element space, providing initial orbit determination from a single detection.

This style of trajectory-match approach, has several advantages beyond just maximising sensitivity to motion models. Coupling measurement parameters together through a trajectory model can improve achievable resolution compared with using separate independent measurement parameters. As an example, a radar's range resolution is determined solely by the signal bandwidth, but its Doppler and Doppler-rate resolution improve with the CPI length.

Through coupling the measurement parameters with the trajectory model, as a radar can resolve finer Doppler and Doppler rate measurements it can essentially resolve finer trajectory states. This can potentially improve target localisation as increasingly accurate state measurements could localise a target within a single range bin.
\vspace{-0.2ex}

\subsection{Zero Doppler Crossing}
\label{ssec:odbd_zdc}
The flexibility of the geometric formulation in Section~\ref{ssec:odbd_search_constraints} allows radar parameters to be used alongside, and in place of, other orbital parameters to constrain the search space. A Doppler shift $f_D$ will define another plane in $\dot{\boldsymbol{r}}$ space, given by:
\begin{equation}
    \label{eq:doppler_plane}
    \frac{\boldsymbol{\rho}}{\rho}\cdot\boldsymbol{\dot{r}} = -\frac{\lambda f_D}{2} + \frac{\boldsymbol{\rho}\cdot\boldsymbol{\dot{q}}}{\rho}~.
\end{equation}
Equation~\eqref{eq:doppler_plane} can be used to search for a particular Doppler shift instead of one of the orbital parameters. This is useful because it allows a blind search to constrain the search-space solely for objects in orbit at their point of closest approach to the sensor. As an object is passing overhead, its point of closest approach will correspond exactly with it being at zero Doppler, which is when it is most detectable\footnote{This may not necessarily hold in all instances, depending on particular beampattern and radar cross section factors.}. If a radar is unable to detect an object at its point of closest approach, at its minimum range, there is little value trying to detect it as it moves further away, towards the horizon.

Another benefit to applying this constraint is that, as Doppler is proportional to the range-rate, this constraint will also restrict the orbit search-space to a point of minimal (or zero) range migration, which greatly simplifies matched-processing\footnote{Depending on the CPI length, it may be possible to make ${\rho}(t) \approx {\rho}$.}.

The vast majority of the objects in an Earth-captured orbit are in a circular, or near-circular, orbit. Searching solely for objects in a circular orbit greatly decreases the potential orbital search space. A circular orbit means the eccentricity of the orbital ellipse is zero, $e=0$, and so \eqref{eq:apogiee_and_perigee_ranges} becomes $a=\lvert\boldsymbol{r}\rvert$. In a circular orbit, the position and velocity vectors will always be perpendicular, so \eqref{eq:parallel_planes} simplifies to $\boldsymbol{r}\cdot\boldsymbol{\dot{r}} = 0$, a single plane instead of two parallel planes. The result is that a three-parameter search, within a region of interest, provides sufficient information to match the closest approach of objects in a circular orbit. For a given position
in a search-region, there will be at most two possible orbits to match against (determined from the intersection of \eqref{eq:vis_viva}, \eqref{eq:parallel_planes}, and \eqref{eq:doppler_plane}). This type of search approach, attempting uncued detection of the most common types of orbit when they are most detectable, is a far more realisable and practical approach than a completely unbounded search through measurement parameters. Additionally, for an eccentric orbit, the orbital velocity and position are perpendicular at perigee \cite{Vallado2001fundamentals}. For typical radar detection ranges, an object in a highly eccentric orbit is likely to be within a radar's field of regard solely at, or near, perigee. Because of this, the same simplification of $\boldsymbol{r}\cdot\boldsymbol{\dot{r}} = 0$ could be used to reduce the number of potential orbits.
\vspace{-1.1ex}
\subsection{Single Channel Orbit Detection}
\label{ssec:odbd_single_channel}
Coupling together measurement parameters is not necessarily new; however, incorporating such techniques into the detection stage offers some significant advantages. By coupling together the measurement parameters using these ODBD methods, it is possible to apply this matched filtering to single beam radar systems. This could be a post-beamformed surveillance signal from an array or even a classic narrowbeam tracking radar. Because the trajectory model determines all measurement parameters, a particular polynomial phase signal which results in a detection is coupled to a particular location and orbit. This is shown in 
\eqref{eq:full_OD_right_here}.  The beamforming parameters do not determine the location; rather the (hypothesised) location determines the beamforming parameters. Removing the array processing, as in \eqref{eq:single_channel}, does not remove the ability to localise a target using the algorithm.

\vspace{-3ex}
\begin{IEEEeqnarray}{rCL}
    \label{eq:single_channel}
    \chi(\boldsymbol{r},\boldsymbol{\dot{r}}) & = \hspace{-1.5ex} \int\limits_{-\frac{T}{2}}^{\frac{T}{2}} & \hspace{-1ex} s(t){d}^*(t-2c^{-1}\rho(\boldsymbol{r},\boldsymbol{\dot{r}},t)){\me}^{-j\frac{2\pi}{\lambda}\dot{\rho}(\boldsymbol{r},\boldsymbol{\dot{r}},t)t}\,dt~~~~
\end{IEEEeqnarray}

In the case of a narrow beam radar, the pointing of the beam will be incorporated into the algorithm by determining the search region that is used. Because it handles sensor motion, this type of processing would be ideal for a satellite-based sensor, with the sensor location term $\boldsymbol{q}(t)$ (or its instantaneous components $\boldsymbol{q}, \boldsymbol{\dot{q}, \boldsymbol{\ddot{q}}}$, etc.) themselves determined by a known orbit rather than the motion of the Earth.

\vspace{-0.2ex}
\section{Simulated Results} \label{sec:results}
These methods have been verified by comparing ODBD-derived measurement parameters, described in section \ref{sec:odbd}, of an object in orbit, against measurement parameters propagated from available ephemerides. These ephemeris tracks consist of the six Keplerian orbital elements, as well as several additional parameters describing drag and orbital decay. These tracks are propagated with the standard SGP-4 propagator used by the USSPACECOM two-line element sets \cite{USSTRATCOM}.

The configuration used for these simulations, matching \cite{8835821}, is a sensor located at the MWA (at a latitude of 27\degree~south) in a bistatic configuration with a  transmitter in Perth, approximately 600 km further south. This transmitter is taken to be transmitting an FM radio signal at a centre frequency of 100 MHz.

\begin{figure}[ht!]
\begin{center}
\includegraphics[width=\columnwidth]{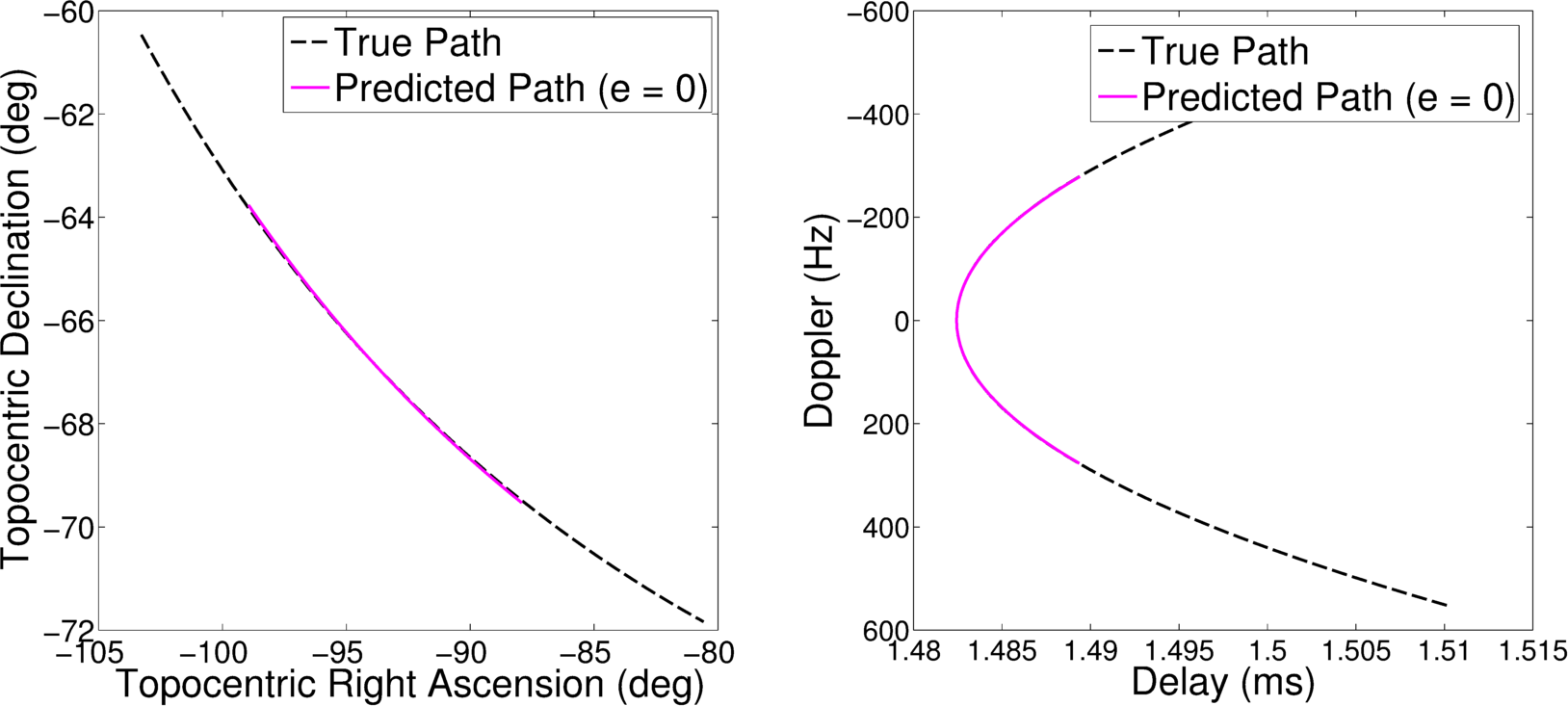}
\end{center}
\vspace{-2ex}
\caption{The measurement parameters of a close pass of an object in a near-circular orbit ($e=0.0007$), as well as the simulation made assuming zero eccentricity at point of closest approach. The left plot is angular space and the right is the delay-Doppler. Twenty seconds of the true pass is shown with ten seconds of the simulated path overlaid.}
\label{fig:standard_circular}
\end{figure}

\begin{figure}[ht!]
\begin{center}
\includegraphics[width=\columnwidth]{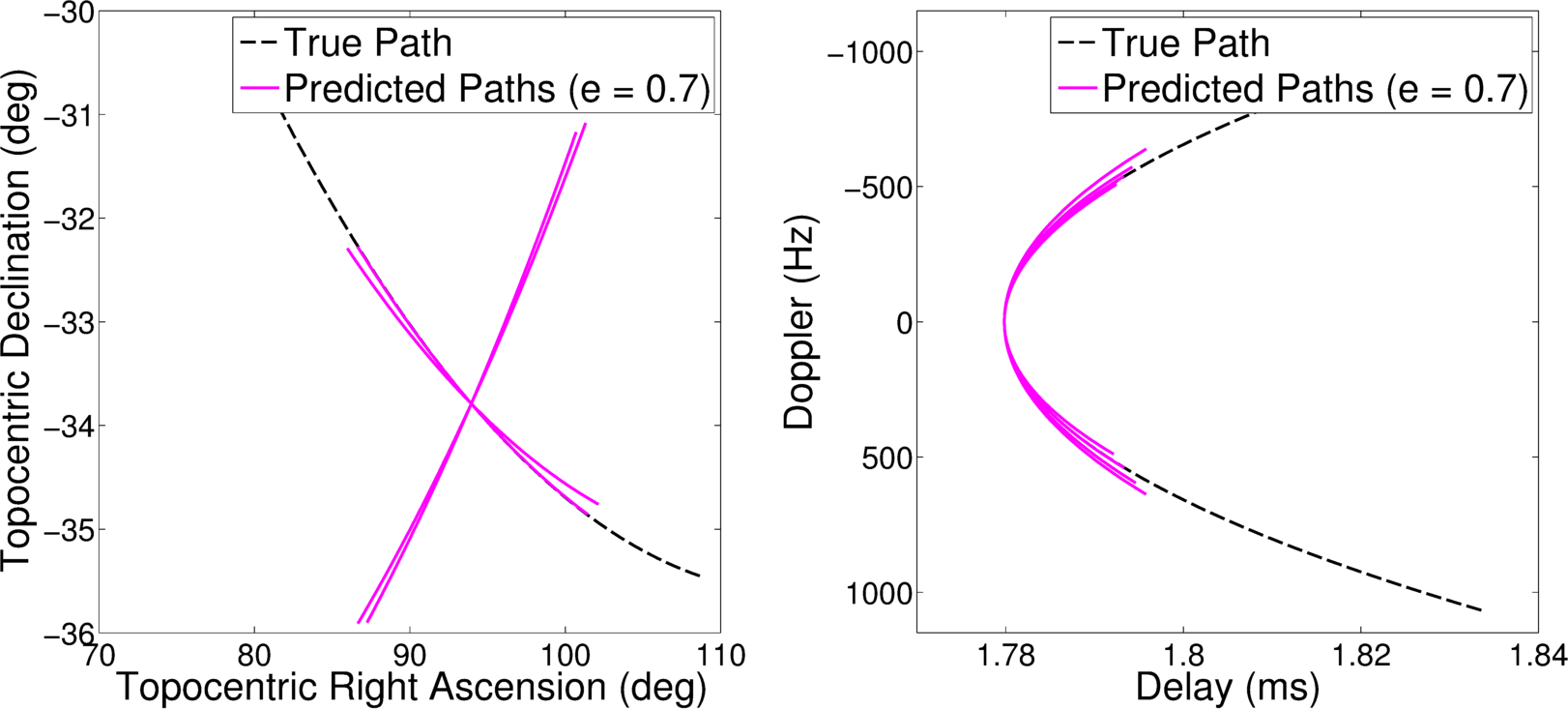}
\end{center}
\vspace{-2ex}
\caption{The measurement parameters of a close pass of an object in an eccentric orbit ($e=0.7$), as well as the four simulations made with the correct eccentricity and semi-major axis. The left plot is angular space and the right is the delay-Doppler. Twenty seconds of the true pass is shown with ten seconds of the simulated paths overlaid.}
\label{fig:standard-eccentric}
\end{figure}

Figure \ref{fig:standard_circular} shows the path of an object in a near circular orbit at closest approach. The simulated measurement parameters match very well in both angular and delay-Doppler space despite being based on a perfectly circular orbit. Likewise, Figure \ref{fig:standard-eccentric} also matches with the prediction, noting that the simulation used the matching eccentricity and semi-major axis.

\begin{figure}[ht!]
\begin{center}
\includegraphics[width=\columnwidth]{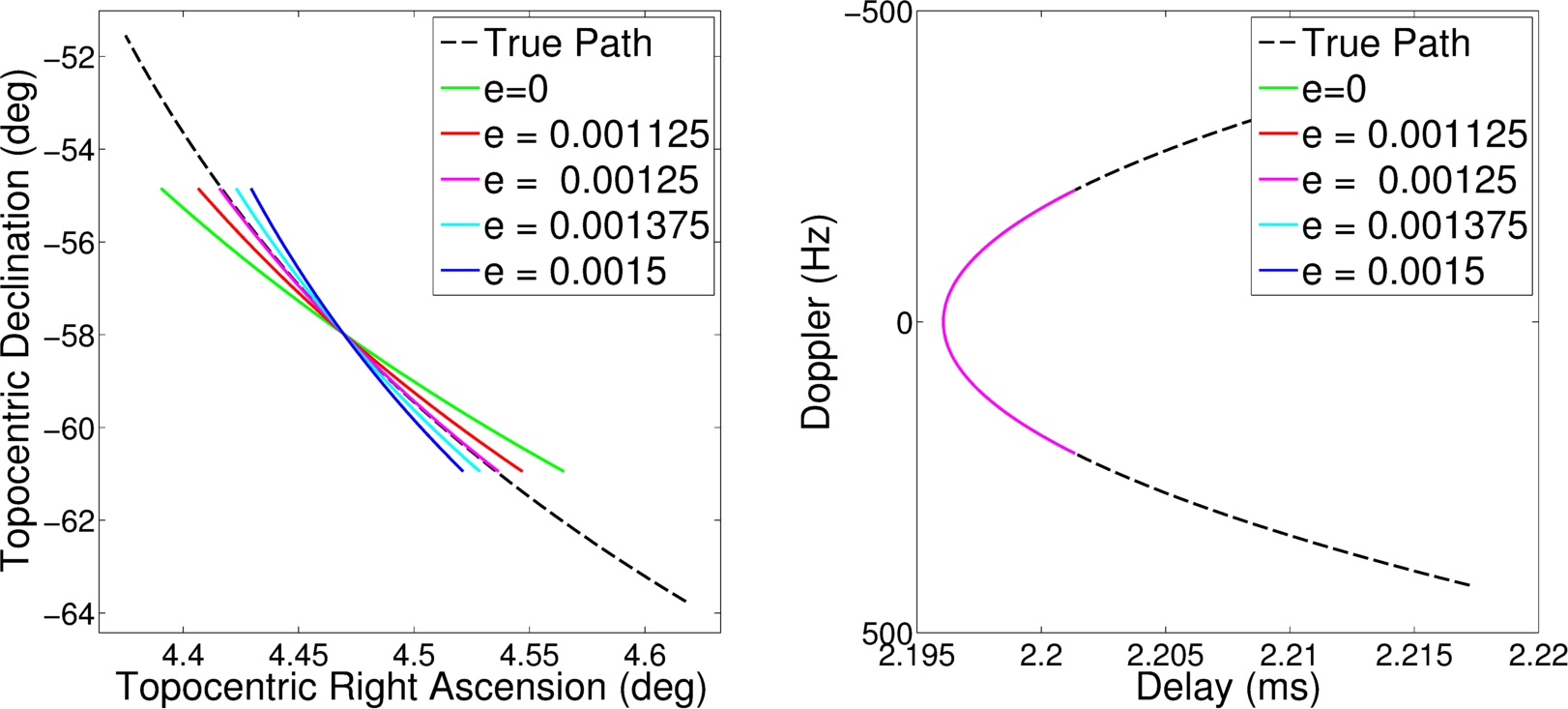}
\end{center}
\vspace{-2ex}
\caption{The measurement parameters of a close pass of an object in a near-circular orbit ($e=0.00126$), as well as several simulations made using different eccentricities. The left plot is angular space and the right is the delay-Doppler. Twenty seconds of the true pass is shown with ten seconds of the simulated paths overlaid.}
\label{fig:eccentricity-mismatch}
\end{figure}

Figure \ref{fig:eccentricity-mismatch} shows the path of an object in a near circular orbit, but slightly more eccentric than Figure \ref{fig:standard_circular}  ($e=0.00126$) at point of closest approach. The simulated circular path matches well in the delay-Doppler space but diverges in the angular space. Additionally, several other simulated close eccentricities are shown, resulting in changes to the direction of travel but little difference in the delay-Doppler space. The delay-Doppler results suggest good tolerance to small eccentricity changes, however the sensor's angular resolution may limit potential processing intervals.

\vspace{-2ex}
\vspace{-2ex}

The good agreement between the parameters derived from methods described in this paper, when compared with ephemeris derived parameters, suggests that earlier results, \cite{8835821}, can be practically achieved without requiring apriori information.

\section{Conclusion} \label{sec:conclusion}
Modern radars are able form matched-filter products with significant numbers of measurement parameters, especially with digital beamforming and extended processing intervals. Conversely, the motion of an object in a Keplerian orbit is defined by only six parameters. Mapping radar measurement parameters from orbital motion parameters constrains the search space for uncued detection, it additionally allows for other constraints to be applied to further reduce the search-space, most notably when searching for objects in a circular orbit at their point of closest approach to the sensor. For a hypothesised orbit of this type, all range, Doppler, and angular motion parameters can be derived entirely from a three-dimensional position.
Detections from this matched filter will correspond to the hypothesised orbit. This means that initial orbit determination can be potentially achieved from a single radar detection.

In future work, these algorithms will be experimentally validated with MWA observations. Noting that although these methods have been developed for the MWA, these methods also apply to conventional active space surveillance radar or even to satellite-based sensors. Additionally, it is planned to investigate the sensitivity of these techniques, characterising their variance by calculating the Cram\'er-Rao lower bound (CRLB) on the variance of the initial orbital estimates.

\bibliographystyle{IEEEtran}
\bibliography{mainbib}

\end{document}